# Electronic Reconstruction Towards Topological Superconductivity in FeTe

Hongtao Rong[1,5], Yang Ge[2,3,5], Zi-Jie Yan[1,5], Haoran Lin[4,5], Bing Xia[1], Xiaoda Liu[1], Zihao Wang[1], Pu Xiao[1], Lok-Kan Lai[1], Stephen Paplini[1], Jiatao Song[1], Jiangang Yang[4], Peter J. Hirschfeld[2], Shuolong Yang[4], Jiabin Yu[2,3], and Cui-Zu Chang[1]

[1]Department of Physics, The Pennsylvania State University, University Park, PA 16802, USA

[2]Department of Physics, University of Florida, Gainesville, FL 32611, USA

[3]Quantum Theory Project, University of Florida, Gainesville, FL 32611, USA

[4]Pritzker School of Molecular Engineering, The University of Chicago, Chicago, IL 60637, USA

[5]These authors contributed equally: Hongtao Rong, Yang Ge, Zi-Jie Yan, and Haoran Lin

Corresponding authors: yangsl@uchicago.edu (S. Y.); yujiabin@ufl.edu (J. Yu); cxc955@psu.edu (C.-Z. C.).

**Abstract: The recent discovery of intrinsic superconductivity in stoichiometric FeTe films[1] has renewed interest in the Te-rich end member of the iron chalcogenides for studies of unconventional and topological superconductivity, yet its intrinsic electronic structure remains unresolved. In this work, we combine molecular beam epitaxy, angle-resolved photoemission spectroscopy (ARPES), electrical transport measurements, density functional theory, and embedded dynamical mean-field theory to track the electronic reconstruction of 20-unit-cell FeTe films as Te annealing progressively removes excess interstitial Fe and drives the system from an antiferromagnetic metal to a superconductor. We find that this evolution is accompanied by recovered quasiparticle coherence, reduced electronic correlations, a Lifshitz transition, and a topological phase transition, yielding $d_{xy}$-dominated hole and electron pockets that favor inter-pocket scattering. In addition, a shallow $d_{xz}/d_{yz}$-derived hole band located about 2 meV below the Fermi level may provide an incipient-band pairing**

**channel, while scattering between the two electron pockets at $\bar{\mathrm{M}}$ may offer additional pairing channels. High-resolution polarization-dependent laser ARPES measurements further reveal a topological surface state whose circular dichroism is consistent with the expected orbital-angular-momentum texture of stoichiometric FeTe. These results establish the intrinsic low-energy electronic structure of superconducting FeTe and identify the electronic states most relevant to superconductivity. The coexistence of intrinsic superconductivity and a topological surface state establishes stoichiometric FeTe as a promising platform for exploring topological superconductivity and possible Majorana bound states[2-6].**

**Main text:** Iron-based superconductors (FeSCs) have provided a versatile platform for investigating the interplay among electronic correlations, magnetism, superconductivity, and band topology[3,7-10] since the discovery of 26 K superconductivity in $LaFeAsO_{1-x}F_x$ in 2008 (ref.[11]). The superconducting transition temperature ($T_c$) quickly reached ≈55 K in the 1111 family[12,13], establishing FeSCs as a major class of unconventional superconductors. The subsequent discovery of superconductivity above 60 K in monolayer FeSe grown on $SrTiO_3$(100) substrates further stimulated interest in iron-chalcogenide superconductors[14]. Among them, the $FeTe_{1-x}Se_x$ family is particularly attractive because its simple PbO-type crystal structure allows systematic tuning of its electronic and magnetic properties through isovalent substitution without introducing additional charge carriers[15,16]. Superconductivity emerges as Se substitutes for Te (refs.[17,18]), while the superconducting transition temperature $T_c$ can be further enhanced under pressure[19]. In parallel, the observation of topological surface states (TSSs) in the $FeTe_{1-x}Se_x$ family[3,20-23] has established these materials as a promising platform for the coexistence of unconventional superconductivity and nontrivial band topology, thereby opening a pathway towards realizing topological superconductivity[2-6].

As the Te-rich end member of this family, FeTe has historically been regarded as a nonsuperconducting bicollinear antiferromagnet[24-26], with a magnetic ordering vector $Q = (\frac{\pi}{2}, \frac{\pi}{2})$ distinct from the collinear antiferromagnetic order, i.e., $Q = (\pi, 0)$, found in many other FeSCs (refs.[27,28]). Early angle-resolved photoemission spectroscopy (ARPES) measurements of $Fe_{1+x}Te$ revealed a nearly electron-hole-compensated Fermi surface (FS) without obvious nesting at the magnetic ordering vector, suggesting that its magnetism cannot be understood within a simple weak-coupling nesting picture[29]. Subsequent ARPES studies, however, reported substantially different FS topologies, low-energy dispersions, and spectral-weight distributions[30-33], highlighting the strong influence of excess interstitial Fe and leaving the intrinsic electronic structure of FeTe unresolved. More recently, post-growth annealing under a Te flux was shown to remove interstitial Fe and to induce intrinsic superconductivity in stoichiometric FeTe (refs.[1,34]), establishing interstitial Fe as a key tuning parameter for its ground state. This discovery raises fundamental questions about how the electronic structure evolves towards stoichiometry, which electronic states accompany superconductivity, and whether stoichiometric FeTe itself hosts nontrivial TSSs.

In this work, we establish the electronic reconstruction that accompanies the transformation from antiferromagnetic $Fe_{1+x}Te$ to superconducting stoichiometric FeTe. As excess interstitial Fe is progressively removed, quasiparticle coherence is restored, electronic correlations are reduced, the FS undergoes a Lifshitz transition[35], and a topological phase transition occurs, concomitant with the emergence of superconductivity. In stoichiometric FeTe, we identify a $d_{xy}$-dominated hole pocket at $\bar{\Gamma}$ and an electron pocket with the same orbital character at $\bar{\mathrm{M}}$, providing favorable conditions for inter-pocket scattering. All $d$ orbitals discussed in this work refer to Fe 3$d$ orbitals. In addition, a shallow $d_{xz}/d_{yz}$-derived hole band remains only ≈2 meV below the Fermi level $E_F$ at $\bar{\Gamma}$, while a $d_{xz}/d_{yz}$-derived electron pocket crosses $E_F$ at $\bar{M}$. Therefore, this shallow hole band may act as an incipient band, providing an additional pairing channel[36-38]. The electron pockets at $\bar{M}$ are well nested despite their distinct orbital characters, providing additional potential pairing channels. High-resolution laser ARPES further resolves a Dirac-cone-like surface state, while circular-dichroism ARPES reveals a characteristic dichroic response consistent with the calculated orbital-angular-momentum texture. Our findings establish the intrinsic low-energy electronic structure of stoichiometric FeTe and position it as a promising platform for investigating topological superconductivity and possible Majorana bound states.

To uncover this evolution, we grow 20-unit-cell (UC) FeTe films by molecular beam epitaxy (MBE) and progressively remove excess interstitial Fe via sequential in situ Te annealing, allowing us to track the same FeTe film continuously from the as-grown antiferromagnetic state to the stoichiometric superconducting state. The MBE growth and annealing processes are monitored by reflection high-energy electron diffraction (RHEED) (Fig. S1) and scanning tunneling microscopy (STM) (Fig. S2), while the electronic structure after each annealing step is measured in situ by helium-lamp ARPES directly connected to the MBE chamber. Electrical transport measurements independently track the evolution of magnetism, superconductivity, and carrier type. To resolve the TSS, we transfer stoichiometric FeTe films under ultrahigh vacuum for high-resolution laser-based ARPES measurements with linear and circular polarizations. We further combine these experiments with density functional theory (DFT) and embedded dynamical mean-field theory (eDMFT) to determine orbital characters, identify a TSS, and capture the strong-correlation effects governing the bulk electronic structure.

**Electronic reconstruction from as-grown to stoichiometric FeTe**

Our recent study[1] demonstrated that Te annealing transforms as-grown FeTe into stoichiometric FeTe by progressively removing excess interstitial Fe through reaction with Te (Fig. 1a). Electrical transport measurements show that the as-grown 20-UC FeTe film is nonsuperconducting and exhibits antiferromagnetic order with a *Néel* temperature $T_{\mathrm{N}} \approx 58$ K, as indicated by the hump in $R_{xx}(T)$ (Figs. 1c and S3a). After subsequent Te annealing, superconductivity emerges. In stoichiometric 20-UC FeTe, the $T_{\mathrm{N}}$ hump becomes indiscernible, while a superconducting transition emerges with an onset temperature $T_{\mathrm{c,onset}} \approx 13.3$ K and zero-resistance temperature $T_{\mathrm{c,0}} \approx 12.2$ K (Figs. 1c and S3c). To track the corresponding evolution of the electronic structure, we divide the Te-annealing treatment into multiple steps and perform ARPES measurements after each in situ Te-annealing step. These measurement cycles are labeled #1-#8, with Cycles #1 and #8 corresponding to the as-grown and stoichiometric 20-UC FeTe films, respectively. We discuss the full electronic evolution below (Figs. 2, 3, and S7-S12). Measurements on multiple samples are highly reproducible. We first compare the FSs and band structures of as-grown and stoichiometric 20-UC FeTe films (Fig. 1). We define the high-symmetry points with respect to the 2-Fe Brillouin zone (BZ) (Fig. 1b, d, i). For the as-grown 20-UC FeTe film, the FS consists of an electron pocket at $\bar{\Gamma}$ (labeled $B_4$), two nearly overlapping electron pockets at $\bar{\mathrm{M}}$ (labeled $M_3$ and $M_4$), and finite spectral weight near $\bar{\mathrm{X}}$ (Figs. 1d and S13a). After Te annealing, the FS of the stoichiometric 20-UC FeTe film exhibits a hole pocket (labeled $B_1$) and a weakly discernible electron pocket (labeled $S_1$) at $\bar{\Gamma}$, together with two nearly overlapping electron pockets at $\bar{\mathrm{M}}$ (i.e., $M_3$ and $M_4$). No band crosses $E_{\mathrm{F}}$ near $\bar{\mathrm{X}}$ (Figs. 1i, S13, and S14). The absence of the pocket near $\bar{\mathrm{X}}$ suggests that stoichiometric FeTe does not enter the orbital-selective Mott phase reported in a prior study[39]. The FS evolution of the FeTe films across Cycles #1-#8 is summarized in Fig. S7.

Band dispersions of the as-grown and stoichiometric FeTe films measured along the $\bar{\mathrm{M}} - \bar{\Gamma} -$

$\overline{\mathrm{M}}$ direction are shown in Fig. 1e,j, respectively. Note that all $\overline{\Gamma}-\overline{\mathrm{M}}$ or $\Gamma-\mathrm{M}$ paths used in this work refer to the horizontal direction indicated by the red arrows in Fig. 1d,i. In the antiferromagnetic state, the as-grown FeTe film exhibits broad spectral features (Fig. 1e). At $\overline{\Gamma}$, an electron-like band $B_4$ is observed, together with a nearly vertical band (labeled $B_3$) at higher binding energy, forming a fork-like feature consistent with prior ARPES studies[31,32]. In addition, two weak hole-like bands (labeled $B_1$ and $B_2$) are observed below $E_F$. These experimental features are well reproduced by our eDMFT calculations with an electron doping of 0.24 electrons/Fe, estimated from the measured Fermi pockets using the Luttinger sum rule[40,41] (Figs. 1f, S15, and S16). In contrast, the low-energy electronic structure of stoichiometric FeTe is markedly different (Fig. 1j). At $\overline{\Gamma}$, two well-defined hole-like bands, $B_1$ and $B_2$, are clearly resolved, with $B_1$ crossing $E_F$. A Dirac-cone-like band $S_1$ is discernible, together with a hole-like band $B_3$ at higher binding energy. These features resemble the electronic structure reported for $FeTe_{1-x}Se_x$ (refs.[42,43]). Except for $S_1$, which is a surface state discussed below, the remaining features match our eDMFT calculations for stoichiometric FeTe (Figs. 1k and S17-S19). Because the $d_{xz}$ and $d_{yz}$ orbitals are related by fourfold screw-rotation symmetry and interchange between orthogonal $\overline{\Gamma}-\overline{\mathrm{M}}$ paths, we denote this orbital character as $d_{xz}/d_{yz}$ unless referring to a specific path. The calculations also show that the $B_4$ band lies above $E_F$ and is derived primarily from the $d_{xy}$ and Te $p_z$ orbitals. The $B_1$ and $B_2$ bands are predominantly derived from the $d_{xy}$ and $d_{xz}/d_{yz}$ orbitals, respectively, whereas the $B_3$ band exhibits mixed $d_{xz}/d_{yz}$ and Te $p_z$ orbital character. The calculated bands span a narrower energy scale near $E_F$ than observed experimentally. Additional weak spectral features (labeled W) are observed in both as-grown and stoichiometric FeTe (Fig. 1e,j) and are reproduced by the corresponding eDMFT calculations (Fig. 1f,k).

Along the $\overline{\Gamma}-\overline{\mathrm{M}}-\overline{\Gamma}$ direction, our eDMFT calculations for as-grown FeTe predict two broad, nearly overlapping electron-like bands, $M_3$ and $M_4$, crossing $E_F$ (Figs. 1h and S16). Two relatively flat bands, $M_1$ (electron-like) and $M_2$ (hole-like), are also present at higher binding energy. Experimentally, only broad and weak electron-like features corresponding to $M_3$ and $M_4$ are discernible in as-grown 20-UC FeTe films (Fig. 1g). After Te annealing, the electronic structure near $\bar{M}$ becomes much better resolved in stoichiometric 20-UC FeTe (Fig. 1l). Our eDMFT calculations predict three electron-like bands, $M_1$, which evolves from $B_1$ at Γ, $M_3$, and $M_4$, of which $M_3$ and $M_4$ cross $E_F$ and form electron pockets smaller than those in as-grown FeTe, together with a hole-like band $M_2$ near $E_F$ (Fig. 1m). The orbital-projected band structures show that $M_1$ and $M_3$ are predominantly of $d_{xy}$ orbital character, whereas $M_2$ and $M_4$ are mainly derived from the $d_{xz}/d_{yz}$ orbitals (Fig. S18). In addition, the incoherent spectral weight (ISW) appears at higher binding energy. In our ARPES spectra (Fig. 1l), the $M_1$ and $M_2$ bands are clearly resolved, whereas the $M_3$ and $M_4$ bands appear as overlapping features, with the ISW clearly visible. At both $\bar{\Gamma}$ and $\bar{M}$, the fitted eDMFT bands overlay the experimental and calculated spectra, showing overall good agreement with experiment despite minor differences in band dispersion. Next, we systematically examine how this electronic structure evolves during Te annealing.

**Band evolution near $\bar{\Gamma}$ during Te annealing**

Figure 2 shows the evolution of the band structure near the $\bar{\Gamma}$ point during Te annealing. For the first three cycles (i.e., #1-#3), the electronic structures remain similar, featuring an electron-like pocket $B_4$, a vertically dispersive band $B_3$ at higher binding energy, and a weak-intensity hole-like band $B_2$ (Fig. 2a). The $B_2$ band is more clearly resolved in the second-derivative spectra (Fig. 2b,c). Parabolic fits to the $B_4$ band reveal a gradual upward shift and steepening of the band as the interstitial Fe concentration decreases (Fig. 2a,b). Consistently, the extracted Fermi momentum $k_F$ decreases progressively from Cycles #1 to #3 (Fig. 2d), in agreement with the evolution of the $B_4$ band. The concurrent increase in the band slope further indicates a reduced effective mass and weaker electronic correlations (Fig. S8). With additional Te annealing, all bands become more pronounced, particularly the $B_2$ band. Energy-distribution curves (EDCs) at $k_{//}$ = -0.2 Å$^{-1}$ (Fig. 2e) show that the spectral weight of the $B_2$ band remains weak during Cycles #1-#3 but increases significantly in Cycle #4, consistent with the band evolution observed in Fig. 2a. For Cycle #4, the fitted $B_4$ band dispersion shifts further upward and becomes steeper, consistent with the further reduction in $k_F$ (Fig. 2d).

From Cycle #5 onward, the band structure changes only slightly. It is characterized by the hole-like $B_2$ and $B_3$ bands and a Dirac-cone-like $S_1$ band (Fig. 2a). Compared with the first four cycles, however, the electronic band structure is distinctly different: the $B_3$ band evolves from a nearly vertical dispersion into a hole-like band, while the electron-like $B_4$ band shifts above $E_F$ and a Dirac-cone-like $S_1$ band emerges. Beginning with Cycle #5, the $B_3$ band splits into two branches, whose momentum separation increases with further Te annealing before approaching a nearly constant value (Fig. S9). Meanwhile, the $B_2$ band spectral weight continues to increase. From Cycle #6 onward, the strong $B_2$-band intensity precludes reliable $S_1$-band fitting. Therefore, we show only the fitted $B_2$-band dispersion. Parabolic fits to the $B_2$ band reveal a gradual upward shift towards $E_F$ with continued Te annealing (Fig. 2a), consistent with the EDCs at $k_{//}$ = -0.05 Å$^{-1}$ (Fig. 2f). Although the EDC peak is associated with the $S_1$ band, its progressive shift towards $E_F$ reflects the overall evolution of the nearby electronic structure. Despite this upward shift, the $B_2$ band maximum remains ≈2 meV below $E_F$ even in Cycle #8, as determined from the fitting results (Fig. S10). In addition, a $d_{xy}$-dominated band, $B_1$, centered at $\bar{\Gamma}$, is observed (Fig. 1j). During Cycles #1-#3, the $B_1$ band is relatively flat and exhibits weak spectral weight (Fig. 2c,g). From Cycle #4 onward, $B_1$ becomes progressively better resolved, with enhanced spectral weight and increased dispersion, indicating weaker electronic correlations. Moreover, from Cycle #6 onward, the $B_1$ band crosses $E_F$ (Figs. 2c and S10). Overall, Te annealing drives a Lifshitz transition[35] at $\bar{\Gamma}$ in the bulk electronic structure, transforming an electron pocket ($B_4$) in as-grown FeTe into a hole pocket ($B_1$) in stoichiometric FeTe.

**Band evolution near $\bar{\mathrm{M}}$ and carrier-type transition during Te annealing**

Next, we examine the evolution of the band structure at the $\bar{\mathrm{M}}$ point (Figs. 3, S11, and S12). During Cycles #1-#3, weak electron-like features corresponding to the $M_3$ and $M_4$ bands are observed near $E_F$ (Fig. 3a), with their relatively flat band bottoms more clearly resolved in the corresponding second-derivative spectra (Fig. 3b). From Cycle #4 onward, the $M_3$ and $M_4$ bands become progressively better defined, accompanied by a gradual increase in spectral weight (Fig. 3a,b). In addition, a weak electron-like band $M_1$ and a hole-like band $M_2$ become visible, and ISW emerges at binding energy $E_B \approx 0.1$ eV below the $M_1$ - $M_4$ bands. The EDCs at $k_{//} = 0$ Å$^{-1}$ show a peak from the $M_1$ - $M_4$ bands, with an additional contribution from ISW (Fig. 3c). These features are weak during Cycles #1-#3 but become progressively enhanced with further Te annealing. Similarly, the EDCs at $k_{//} = -0.63$ Å$^{-1}$ reveal the emergence of the $M_1$ and $M_2$ bands from Cycle #4 (Figs. 3d and S11). Distinct peaks corresponding to these two bands are resolved from Cycle #5 and continue to increase in intensity with further Te annealing. To quantitatively track the evolution of the $M_3$ and $M_4$ bands, we overlay parabolic fits to these bands in Fig. 3a,b. We do not show fits for Cycles #1-#3 because weak spectral intensity prevents reliable fitting. From Cycle #4 onward, the slopes of the $M_3$ and $M_4$ bands gradually increase, indicating reduced effective masses and weaker electronic correlations (Fig. S12). Meanwhile, the $k_F$ decreases from Cycle #1 to Cycle #5 and then remains nearly constant at $\approx$0.27 Å$^{-1}$ (Fig. 3e). Owing to the broad spectral features near $\bar{\mathrm{M}}$, no further changes are clearly resolved during Cycles #6-#8.

Te annealing also induces pronounced changes in electrical transport properties. Hall measurements show that the dominant carrier type evolves from electron-type in the as-grown FeTe films to hole-type in superconducting FeTe films with $T_{c,0} > 9$ K, as indicated by the extracted 2D carrier densities $n_{2D}$ (Figs. 3f and S4). Upon cooling, the hole-type carrier density increases significantly, whereas the electron-type carrier density remains nearly $T$-independent. For superconducting FeTe films with $T_{c,0} < 9$ K, corresponding to Cycles #5 and #6, the extracted $n_{2D}$ approaches zero and exhibits apparent sign reversals, indicating nearly compensated electron and hole contributions to the Hall response (Fig. S5). The evolution of the Hall response can be further understood from the corresponding FS reconstruction. While the electron pockets near $\bar{M}$ persist throughout Te annealing, the FS near $\bar{\Gamma}$ undergoes a Lifshitz transition. As prior studies of FeSe have shown that hole carriers can exhibit longer scattering times and higher mobilities than electron carriers at low $T$ (refs.[44,45]), the upward shift of the $B_4$ electron pocket above $E_F$ and the emergence of the $B_1$ hole pocket likely contribute to the crossover from electron- to hole-dominated Hall transport in our FeTe films.

**TSS in stoichiometric FeTe**

The DFT-calculated band structure of stoichiometric FeTe throughout the full BZ reveals three hole-like bands near $E_F$ at Γ, two of which cross $E_F$ (Figs. 4a,b and S20). In the $k_z = \pi$ plane, only two bands are present at Z, and both cross $E_F$. Four bands are present at M and A. Stoichiometric FeTe exhibits a dispersive δ band that crosses the α and β bands split by spin-orbit coupling (SOC) along the Γ – Z direction (Fig. 4c), similar to $FeTe_{1-x}Se_x$ (refs.[3,20,46]). The δ band is derived from the $d_{xy}$ and Te $p_z$ orbitals and is referred to as the $xy^-$ band[47], which has odd parity at Γ and Z, whereas the α and β bands have mixed $d_{xz}$ and $d_{yz}$ orbital character and even parity at Γ and Z. Crucially, the orbitally distinct δ and α bands undergo a band inversion along Γ – Z. SOC further opens a hybridization gap at the band crossing, giving rise to a TSS, as revealed by the projected (001) surface band structure (Fig. 4d) and a nontrivial Wilson loop winding (Fig. S21b).

To experimentally verify the predicted nontrivial topology, we perform high-resolution laser ARPES measurements using linear polarizations. The experimental geometry is shown in Fig. S22. Matrix-element analysis shows that, along the $\bar{\Gamma}$ - $\bar{\mathrm{X}}$ path considered here, *s*-polarized light preferentially visualizes the $d_{xz}$ orbital contribution with enhanced sensitivity, whereas *p*-polarized light is more sensitive to the $d_{xy}$, $d_{yz}$, and $p_z$ orbitals. The TSS contains contributions from all of these orbitals and is therefore visible under both polarizations. Under *s*-polarized light, both the parabolic bulk band $B_3$ with dominant $d_{xz}$ orbital character and the TSS ($S_1$) are clearly resolved (Figs. 4e and S20). Under *p*-polarized light, the TSS ($S_1$) remains visible, together with the bulk band $B_2$ dominated by the $d_{yz}$ orbital character (Figs. 4f and S20). The absence of the TSS $S_1$ in as-grown FeTe and its emergence after Te annealing can be understood from the eDMFT band structures along the $\Gamma - \mathrm{Z}$ direction (Fig. S19). In as-grown FeTe, the $p_z$-derived $B_3$ band lies entirely below the $d_{xz}/d_{yz}$-derived $B_2$ and $B_4$ bands, precluding band inversion. However, in stoichiometric FeTe, the $xy^-$ band (i.e., $B_4$) crosses the $d_{xz}/d_{yz}$-derived $B_2$ and $B_3$ bands, leading to the emergence of a TSS when SOC opens a gap at the band crossings.

In a topological insulator[48-51], the TSS exhibits a helical spin texture, with the spin polarization locked perpendicular to the momentum and winding by $2\pi$ around $\bar{\Gamma}$. Owing to strong SOC, the corresponding orbital angular momentum (OAM) is also locked to the momentum and parallel to the spin angular momentum (SAM) (Figs. 4g and S21c). This parallel locking between OAM and SAM persists until the surface state merges into the bulk bands outside the gap. To experimentally probe the predicted OAM texture, we further perform laser-based circular-dichroism ARPES measurements. A pronounced dichroic signal is observed for the TSS (Fig. 4h). The circular-dichroism signal changes sign between opposite momenta and reverses again across the Dirac point, consistent with the calculated OAM texture (Fig. 4g). It is noticeable that the underlying bulk states show apparent circular dichroism signals, which can be due to interatomic interference from the local OAM contributions from different atomic sites[52].

**Discussion and outlook**

By combining ARPES with electrical transport measurements, we establish a systematic evolution of the electronic structure in FeTe films as excess interstitial Fe is progressively removed. FeTe films with relatively high interstitial Fe concentrations during Cycles #1-#3 exhibit a low-$T$ upturn in the $R$-$T$ curves and broad, weakly coherent electronic states (Figs. 1c and S3a). Near $\bar{\Gamma}$, the electronic structure consists of an electron-like pocket $B_4$, a nearly vertical dispersive band $B_3$ at higher binding energy, and two weak hole-like bands $B_1$ and $B_2$, while two weak, nearly overlapping electron-like pockets $M_3$ and $M_4$ cross $E_F$ near $\bar{M}$. As the interstitial Fe concentration decreases, the low-$T$ upturn is suppressed (Fig. S3b) and the spectral weight of the $B_1$ and $B_2$ bands at $\bar{\Gamma}$ increases, accompanied by the emergence of the $M_1$ and $M_2$ bands in Cycle #4. Upon further Te annealing, superconductivity emerges in Cycle #5 (Fig. S3b), accompanied by progressively sharper electronic bands, an upward shift of the electron-like $B_4$ band above $E_F$, and the emergence of the TSS ($S_1$).

With the emergence of zero resistance in Cycles #6-#8 (Figs. 1c and S3c), well-defined quasiparticle peaks indicate the recovery of quasiparticle coherence. The reconstructed FS contains a $d_{xy}$-dominated $B_1$ hole pocket and a small electron pocket $S_1$ (i.e., the TSS) at $\bar{\Gamma}$, together with two nearly overlapping electron pockets $M_3$ and $M_4$ at $\bar{M}$, with predominantly $d_{xy}$ and $d_{xz}$ /$d_{yz}$ orbital character, respectively. The emergence of the $B_1$ hole pocket also supports the evolution towards a positive Hall response. The presence of $d_{xy}$-derived Fermi pockets at both $\bar{\Gamma}$ and $\bar{M}$ suggests that recovering $d_{xy}$ quasiparticle coherence may facilitate inter-pocket scattering[53-57]. In Cycle #8, the $d_{xz}/d_{yz}$-derived $B_2$ band remains only ≈2 meV below $E_F$. Given the proximity of the $B_2$ band to $E_F$, it may serve as an incipient band and provide a possible pairing channel with the $M_4$ electron pocket[36-38]. Meanwhile, the electron pockets at $\bar{M}$ provide additional possible inter-pocket scattering channels, though the scattering may be weak due to their different orbital natures. In addition, electronic correlations progressively weaken from as-grown to stoichiometric FeTe. High-resolution laser ARPES measurements using linear polarizations confirm the Dirac-cone-like $S_1$ band, while circular-dichroism ARPES reveals a characteristic helical dichroic response. The agreement between the observed dichroic response and the calculated OAM and SAM textures provides strong evidence for a TSS in stoichiometric FeTe.

To summarize, we demonstrate that the low-energy electronic structure and electronic correlations of FeTe can be continuously tuned through the controlled removal of interstitial Fe, revealing the intrinsic ground state of stoichiometric superconducting FeTe. The emergence of superconductivity is accompanied by recovered quasiparticle coherence, a Lifshitz transition, and a topological phase transition. The $d_{xy}$-dominated Fermi pockets at $\bar{\Gamma}$ and $\bar{M}$ provide a possible inter-pocket scattering channel despite their different sizes and imperfect nesting. Meanwhile, a shallow $d_{xz}/d_{yz}$-derived incipient band at $\bar{\Gamma}$ may provide an additional pairing channel involving the $d_{xz}/d_{yz}$-derived electron pocket at $\bar{M}$. The coexistence of $\bar{\Gamma}$-centered hole pockets and $\bar{M}$-centered electron pockets may favor $s_{\pm}$ pairing[56,58-61]. In addition, electron pockets at $\bar{M}$ provide further potential scattering channels. Their relatively good nesting may favor *d*-wave pairing[58], although the different orbital characters of these pockets may weaken the corresponding inter-pocket scattering[59,60]. Finally, the coexistence of intrinsic superconductivity and an experimentally established TSS makes stoichiometric FeTe a promising platform for exploring topological superconductivity and possible Majorana bound states[2-6].

## Methods

### MBE growth and Te annealing treatments

All FeTe films are synthesized in a commercial MBE chamber (Scienta Omicron) with a vacuum better than $2 \times 10^{-10}$ mbar. Both metallic 0.5% Nb-doped $SrTiO_3$(100) and insulating $SrTiO_3$(100) substrates are first soaked in deionized water at ≈80 °C for 2 h and then immersed in a diluted hydrochloric acid solution (≈4.5% w/w) for 2 h. These $SrTiO_3$(100) substrates are subsequently annealed at ≈974 °C for 3 h in a tube furnace under a flow of high-purity oxygen. After these treatments, the $SrTiO_3$(100) surfaces become passivated and atomically flat, making them suitable for MBE growth of FeTe films. Next, the heat-treated $SrTiO_3$(100) substrates are loaded into the MBE chamber and outgassed at 600 °C for 1 h. High-purity Fe (99.995%) and Te (99.9999%) are evaporated from Knudsen effusion cells. The substrate temperature is maintained at ≈300 ℃ during growth. The growth rate is ≈0.2 UC/min, calibrated by measuring FeTe film thickness using atomic force microscopy.

To remove excess interstitial Fe, we anneal the as-grown FeTe films under a Te flux at 240 ℃. In our experiments, each Te-annealing step lasts 4 min, and we obtain stoichiometric FeTe films after 7 cycles. These 7 cycles correspond to Cycles #2-#8 in the main text. We monitor MBE growth and Te-annealing treatments using RHEED patterns (Fig. S1). For ex situ electrical transport measurements, a 10-nm-thick Te capping layer is deposited on the FeTe films at room temperature before removal from the MBE chamber.

### Helium-lamp ARPES measurements

Helium-lamp ARPES measurements are performed using a Scienta Omicron DA30 electron energy analyzer directly connected to the MBE chamber. After MBE growth and each Te-annealing step, the FeTe films are transferred in situ from the MBE chamber directly to the ARPES chamber. A helium discharge lamp is used as the light source, providing He I photons with $hv$ = 21.218 eV. The energy and angular resolutions are set to ≈10 meV and ≈0.3°, respectively. All helium-lamp ARPES measurements are performed at $T$ = 30 K under ultrahigh vacuum at a base pressure below $5 \times 10^{-11}$ mbar.

**Laser ARPES measurements**

Laser ARPES experiments are performed using a Scienta Omicron DA30 electron energy analyzer at the University of Chicago[61]. We grow stoichiometric FeTe films by MBE at Penn State, transfer them into an ultrahigh-vacuum suitcase with a base pressure below $5 \times 10^{-10}$ mbar, and then transport them to the multi-resolution photoemission spectroscopy platform at the University of Chicago. The spatial and energy resolutions are ≈10 × 15 μm and ≈2.4 meV, respectively, using 206-nm pulses with a repetition rate of 80 MHz. All laser ARPES measurements are performed at $T$ = 25 K under ultrahigh vacuum, with a base pressure below $5 \times 10^{-11}$ mbar.

**Electrical transport measurements**

FeTe films grown on 2 mm × 10 mm heat-treated $SrTiO_3$ (100) substrates are mechanically scratched into a Hall-bar geometry using a computer-controlled motorized probe station. The effective Hall-bar area is ≈1 mm × 0.5 mm. Electrical contacts are formed by pressing indium spheres onto the Hall bar. Electrical transport measurements are performed using a Physical Property Measurement System (PPMS, Quantum Design DynaCool, 1.7 K, 9 T). An excitation current of 1 μA is used for all $R_{xx}$-$T$ measurements. To minimize oxidation, we measure all FeTe films within 30 minutes after removal from the MBE chamber.

**STM measurements**

STM measurements are performed in a Unisoku 1300 system with a base vacuum better than 2 × $10^{-10}$ mbar. The system incorporates a single-shot $^3$He cryostat to achieve a base temperature of ≈310 mK. Before STM measurements on as-grown and stoichiometric 20 UC FeTe films, the polycrystalline PtIr tips are routinely conditioned on an MBE-grown Ag film to ensure clean and stable tunneling characteristics. The setpoints for all STM measurements are provided in the corresponding figure captions. All STM images are processed using WSxM 5.0 software[62].

**Theoretical calculations**

The electronic structure of FeTe is studied using DFT and eDMFT. For standalone DFT calculations, we calculate the Kohn-Sham electronic structure of stoichiometric FeTe, including SOC, using the Vienna Ab initio Simulation Package (VASP) within the projector-augmented-wave framework[63-67]. We treat exchange and correlation using the Perdew-Burke-Ernzerhof generalized-gradient approximation[68]. The lattice constants are fixed to the experimentally determined values: $a = b = 3.862$ Å, $c = 6.262$ Å, with an Fe-Te plane separation of 1.762 Å. The DFT electronic structure is converged using a 15 × 15 × 10 Monkhorst-Pack $k$-point mesh. The converged calculations show that the electronic states near $E_F$ are derived primarily from Fe $d$ and Te $p$ orbitals.

To facilitate further investigation of the electronic structure, including surface states, the DFT bands are subsequently Wannierized using Wannier90 (ref.[69]), with Fe $d$ and Te $p$ orbitals as projectors, to construct an effective tight-binding (TB) model[1,60]. We define the $x$ and $y$ axes along the nearest-neighbor in-plane Fe-Fe directions for labeling the $p$ and $d$ orbitals. The resulting bulk TB band structure agrees well with the DFT result within ±1.40 eV of $E_F$. As shown in Fig. 4b,c, the DFT band structures are plotted as black lines, while the orbital weights are obtained from the TB calculations. The TB model is also used to calculate the spectral weight at the (001) surface of a semi-infinite FeTe sample using a Green's function method[70]. The calculated spectral weight contains contributions from both surface states and bulk states that extend to the surface, enabling direct comparison with ARPES measurements. Figure 4d shows the calculated spectral weight along the $\bar{X} - \bar{\Gamma} - \bar{X}$ direction, which reproduces the surface states observed in our ARPES measurements (Fig. 4e,f). Wilson loop calculations[71] in the $k_y = 0$ plane using the TB model reveal a nontrivial phase winding (Fig. S21b), demonstrating a nontrivial 2D $\mathbb{Z}_2$ topology[72] on this plane and thereby confirming the topological nature of the surface states. However, the $\mathbb{Z}_2$ topology of the 3D bulk is not well-defined because the system is gapless. Using a slab calculation with a (001) surface, we further calculate the expectation value of the atomic OAM for the surface electronic states. Figure 4g shows the in-plane component of the atomic OAM perpendicular to the $\bar{M} - \bar{\Gamma} - \bar{M}$ direction. Two linearly crossing dispersion branches with opposite OAM are observed, consistent with the strong circular-dichroism ARPES signal in our experiments (Fig. 4h). Within the bulk gap, the SAM is aligned with the atomic OAM (Fig. S21c).

DFT does not fully capture the strong correlation effects of the Fe 3*d* orbitals, which dominate the electronic states near $E_F$. In particular, DFT predicts higher band energies at Γ and lower band energies at M than those observed by ARPES. To better reproduce the experimental results, we calculate the bulk electronic structure using the eDMFT method implemented in the DMFT-WIEN2k suite[73-77]. In eDMFT, the strongly correlated orbitals are treated as an interacting impurity problem embedded, within the DMFT framework, in an environment defined by the DFT electronic structure. We treat the double counting of electronic correlations between DFT and DMFT using an exact double-counting scheme[75]. The eDMFT calculations start from a DFT electronic structure (Figs. S15 and S17) obtained using the all-electron augmented-plane-wave-plus-local-orbital method implemented in WIEN2k (ref.[78]). A continuous-time quantum Monte

Carlo impurity solver is then used to calculate the self-energy of the Fe *d* orbitals within a quantum impurity model embedded in the electronic bath obtained from DFT (ref.[79]). The converged DMFT self-energy is subsequently fed back into the DFT calculation as part of the Kohn-Sham Hamiltonian. This cycle is repeated until full self-consistency between the DFT and DMFT calculations is achieved[74].

For the eDMFT calculations of the bulk electronic structure, the overall numerical setup follows ref.[22]. SOC is neglected because it has only a minor effect on the bulk band structure and orbital character of FeTe (ref.[22]). To enable a direct comparison with our ARPES measurements, eDMFT calculations are performed for two types of FeTe films: (*i*) stoichiometric FeTe and (*ii*) as-grown FeTe containing excess interstitial Fe. The effect of interstitial Fe is modeled as electron doping[80-83], with a compensating increase in the background charge to maintain overall charge neutrality. The doping level is estimated from the Luttinger count based on the ARPES-measured FS areas of all three electron pockets, including one at $\bar{\Gamma}$ and two at $\bar{\text{M}}$. No hole pockets are observed in the as-grown FeTe sample. Summing the areas of these electron pockets yields an estimated doping level of 0.24 electrons per Fe. We note that doping goes beyond a simple Fermi-level shift even at the DFT level (Figs. S15 and S17), as it modifies the charge distribution and consequently the Hartree and exchange-correlation contributions to the Kohn-Sham functional. For the DFT calculations in WIEN2k, we use the local-density approximation for the exchange-correlation potential. In the DMFT calculations, we determine the self-energies of the strongly correlated Fe *d* orbitals using a single-impurity solver, since all Fe sites are effectively symmetry-related at low energies. To ensure convergence of the self-energy, 60 million Monte Carlo steps are performed in each DMFT iteration at $T = 116$ K. A fully rotationally invariant on-site Coulomb repulsion is used with $U = 5$ eV and Hund's coupling $J = 0.8$ eV, following established values for iron chalcogenides[84,85]. After achieving eDMFT self-consistency, we analytically continue the self-energy to the real-frequency axis using the maximum entropy method and calculate the electronic spectral weights. The resulting orbital-resolved spectral weights are shown in Figs. 1f,h and S16 for electron-doped FeTe, and in Figs. 1k,m and S18 for FeTe. Compared with DFT, the eDMFT calculations lower the hole-pocket energy at Γ and raise the electron-pocket energy at M, thereby renormalizing both towards $E_F$. The eDMFT spectra also exhibit substantial correlation-induced broadening near $E_F$, in good agreement with our ARPES measurements (Fig. 1).

**Acknowledgments:** We thank N. Dihingia, A. J. Grutter, D. Reifsnyder Hickey, C. X. Liu, N. Samarth, W. Wu, B. Yan, and M. Yi for helpful discussions, and A. Richardella for technical assistance. This work is primarily supported by the DOE grant (DE-SC0023113), including the MBE growth and ARPES measurements. Electrical transport measurements are supported by the ONR grant (N000142412133) and the Penn State MRSEC for Nanoscale Science (DMR-2011839). STM/S measurements are supported by the NSF grant (DMR-2241327). The MBE growth and ARPES measurements are partially performed in the NSF-supported 2DCC MIP facilities (DMR-2039351). Laser-based ARPES measurements are supported by the NSF grant (DMR-2145373). MBE growth and electrical transport measurements performed at UChicago are supported by the NASA Space Technology Research Grants Program (80NSSC25K7832). PJH acknowledges the support from the NSF grant (DMR-2231821) and the Leverhulme Trust. JY acknowledges the support from the University of Florida startup funds. CZC acknowledges the support from the Gordon and Betty Moore Foundation's EPiQS Initiative (GBMF9063 to C.-Z. C).

**Author contributions:** CZC conceived and designed the experiments. HR, ZJY, ZW, PX, LKL, and CZC performed the MBE growth. HR and XL performed the helium-lamp ARPES measurements. HL, JYang, and SY performed the laser-based ARPES measurements. ZJY, HR,

XL, PX, LKL, and CZC conducted the electrical transport measurements. BX, ZW, SP, JS, and CZC performed the STM/S measurements. YG, PJH, and JYu provided theoretical support. HR, ZJY, YG, HL, and CZC analyzed the data and wrote the manuscript with input from all authors.

**Competing interests:** The authors declare no competing interests.

**Data availability:** The data that support the findings of this article are openly available at Zenodo (https://doi.org/10.5281/zenodo.22051910).

## Figures and figure captions:

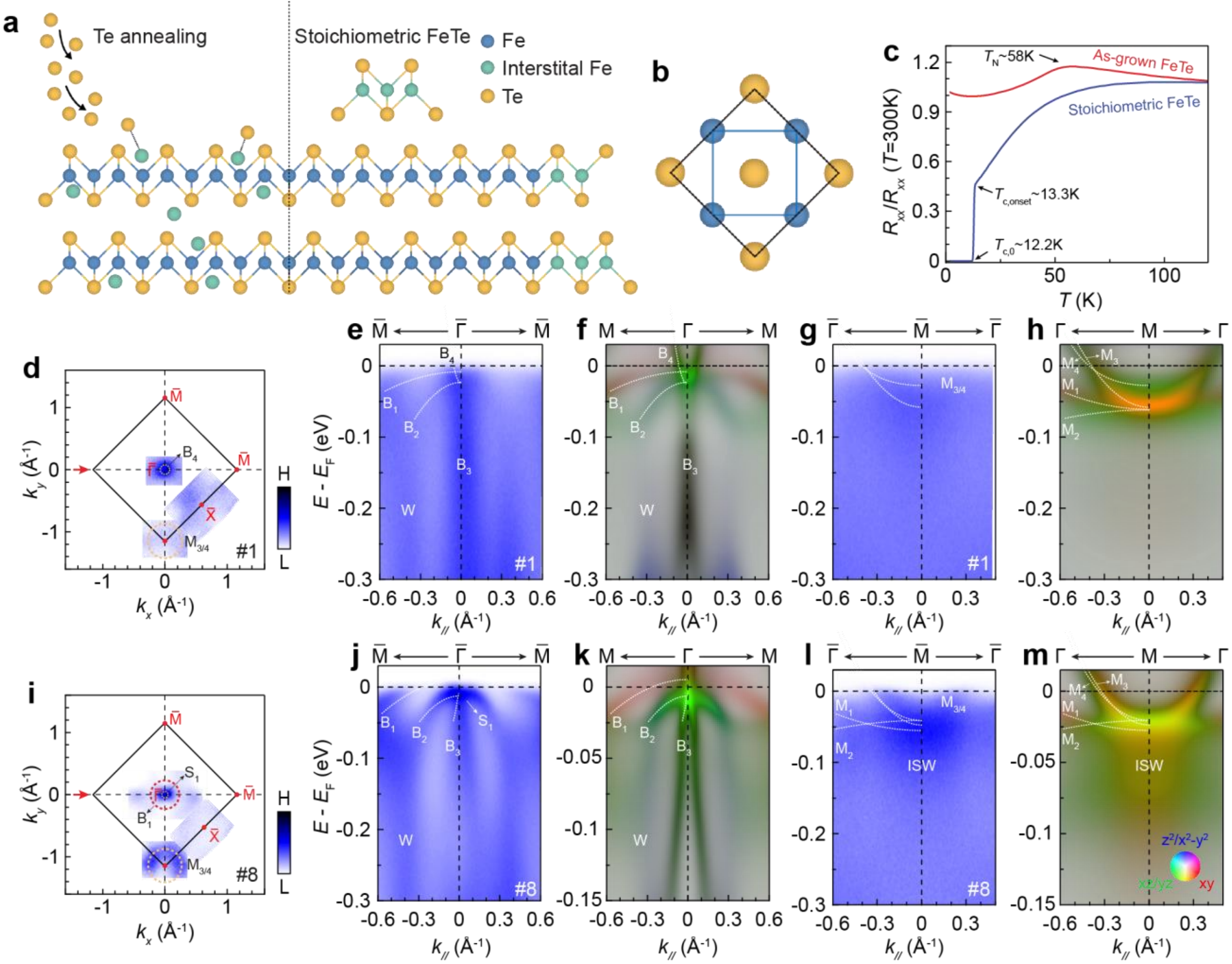

**Fig. 1| Electronic band-structure comparison between as-grown and stoichiometric 20-UC FeTe films. a,** Schematic of the transformation from as-grown to stoichiometric FeTe films through Te annealing. **b,** Top view of the FeTe crystal structure. **c,** $T$-dependent $R_{xx}/R_{xx}(T\text{=300K})$ of as-grown and stoichiometric FeTe films. Black arrows mark $T_{\mathrm{N}}$ of the as-grown FeTe film and $T_{\mathrm{c,onset}}$ and $T_{\mathrm{c,0}}$ of the stoichiometric FeTe film. **d, i,** FS maps of the as-grown (**d**) and stoichiometric (**i**) FeTe films. Red arrows in (**d, i**) indicate the $\bar{\mathrm{M}} - \bar{\Gamma} - \bar{\mathrm{M}}$ and $\mathrm{M} - \Gamma - \mathrm{M}$ paths used in this work. **e, j,** Band dispersions measured along the $\bar{\mathrm{M}} - \bar{\Gamma} - \bar{\mathrm{M}}$ direction for the as-grown (**e**) and stoichiometric (**j**) FeTe films. **f, k,** Calculated band dispersions along the $\mathrm{M} - \Gamma - \mathrm{M}$ direction for the as-grown (**f**) and stoichiometric (**k**) FeTe bulk crystals using eDMFT calculations without SOC at $k_z = 0$, with the Te height fixed at 1.762 Å. The calculation in (**f**) includes an additional electron doping of 0.24 electrons/Fe. **g, l,** Same as (**e, j**), but measured along the $\bar{\Gamma} - \bar{\mathrm{M}} - \bar{\Gamma}$ direction. **h,**

**m,** Same as (**f, k**), but calculated along the $\Gamma - \mathrm{M} - \Gamma$ direction. The calculated bands in (**f, h**) are shifted by 20 meV towards higher binding energy, and those in (**k, m**) by 17.5 meV. The white dashed curves in (**e-h**, **j-m**) fit the local second-derivative minima near each band maximum, using energy derivatives for all bands except $B_4$ and momentum derivatives for $B_4$.

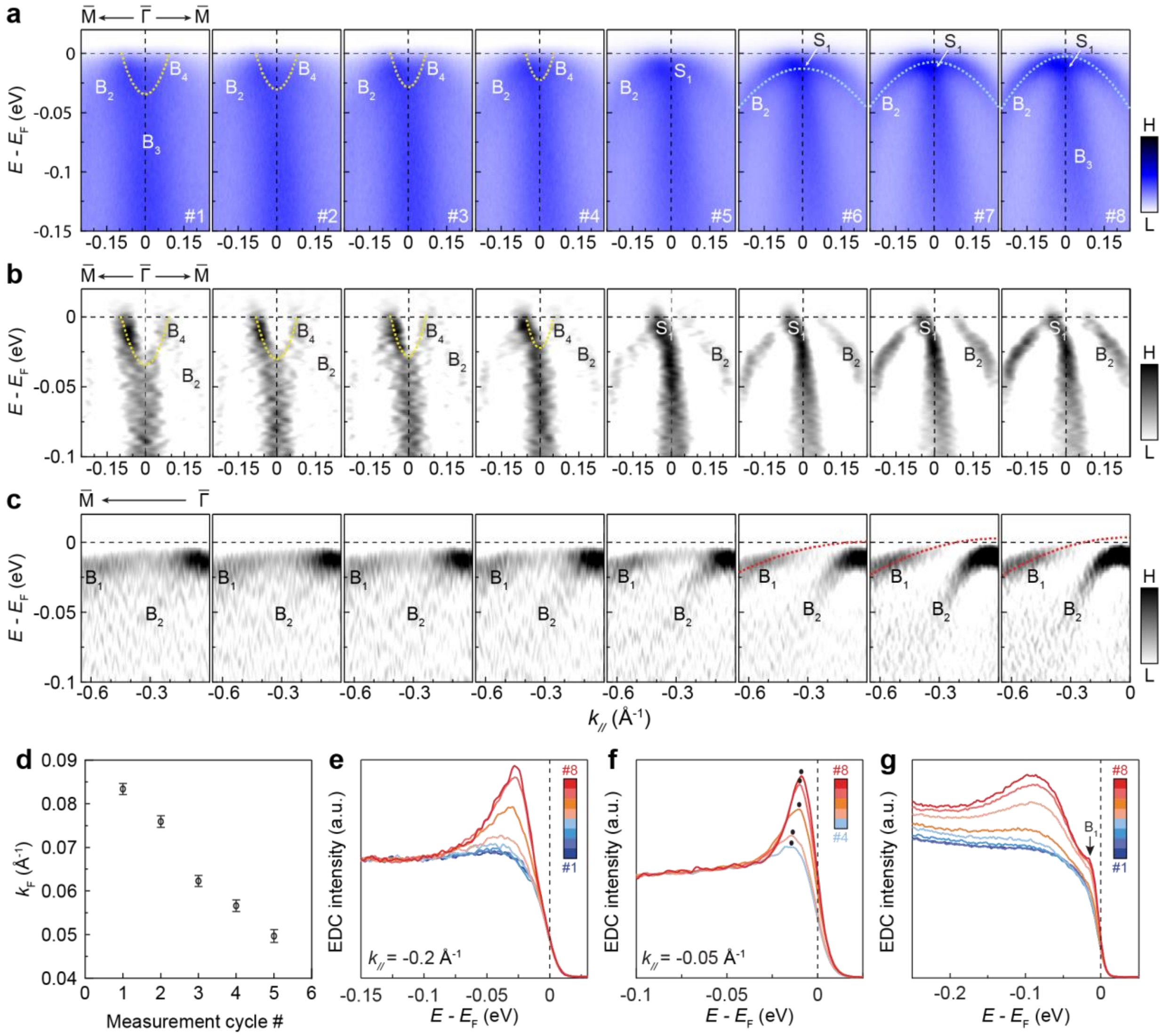

**Fig. 2| Te-annealing-induced band evolution near the $\overline{\Gamma}$ point in the 20-UC FeTe film. a,** Te-annealing-induced band evolution measured along the $\overline{\mathrm{M}}-\overline{\Gamma}-\overline{\mathrm{M}}$ direction. **b,** Corresponding second-derivative spectra with respect to momentum over the energy range from -0.1 eV to 0.025 eV. **c,** Te-annealing-induced band evolution in second-derivative spectra with respect to energy along the $\overline{\mathrm{M}}-\overline{\Gamma}$ direction. Parabolic fits to the $B_4$ band are overlaid in (**a**, **b**). Parabolic fits to the $B_2$ band are overlaid in (**a**), and parabolic fits to the $B_1$ band are overlaid in (**c**). **d,** Extracted Fermi momenta $k_F$ for Cycles #1-#5. **e,** EDCs at $k_{//}$ = -0.2 Å$^{-1}$ for Cycles #1-#8. **f,** EDCs at $k_{//}$ = -0.05 Å$^{-1}$ for Cycles #4-#8. The peak position of each EDC is marked by a black solid circle. **g,** Integrated EDCs obtained by integrating over the momentum range -0.525 Å$^{-1}$ < $k_{//}$ < -0.275 Å$^{-1}$ for Cycles #1-#8.

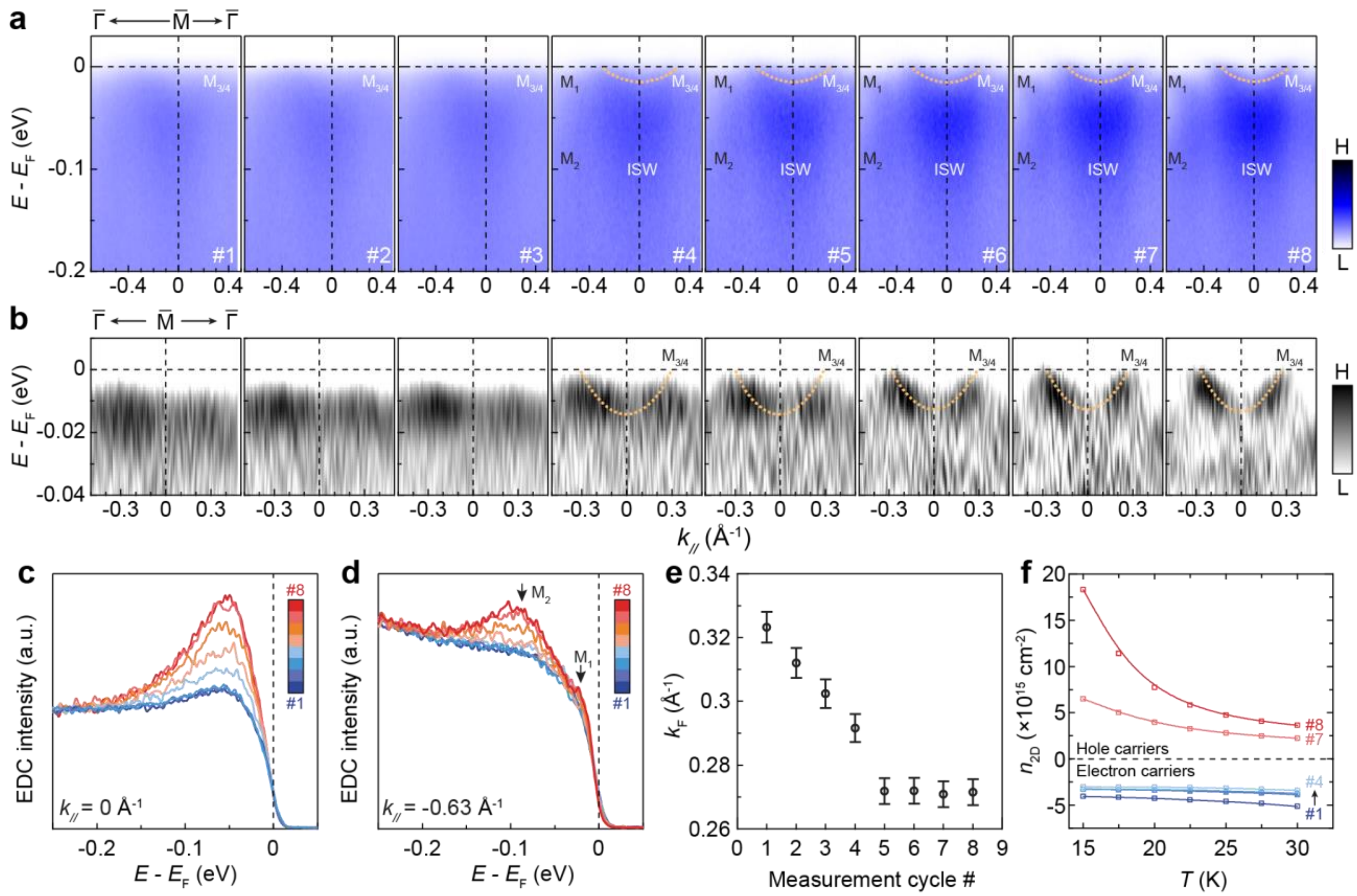

**Fig. 3| Te-annealing-induced band evolution near the $\overline{\mathrm{M}}$ point in the 20-UC FeTe film. a,** Te-annealing-induced band evolution measured along the $\overline{\Gamma} - \overline{\mathrm{M}} - \overline{\Gamma}$ direction. **b,** Corresponding second-derivative spectra with respect to energy over the energy range from -0.04 eV to 0.01 eV and the momentum range from -0.5 $Å^{-1}$ to 0.5 $Å^{-1}$. Parabolic fits to the $M_{3/4}$ bands are overlaid in (**a**, **b**). **c, d**, EDCs at $k_{//} = 0$ $Å^{-1}$ (**c**) and $k_{//} = -0.63$ $Å^{-1}$ (**d**) for Cycles #1-#8. The positions of the $M_1$ and $M_2$ bands are marked by black arrows in (**d**). **e,** Extracted Fermi momenta $k_F$ for Cycles #1-#8. **f,** $T$ dependence of the 2D carrier density $n_{2D}$ for Cycles #1-#4, #7, and #8.

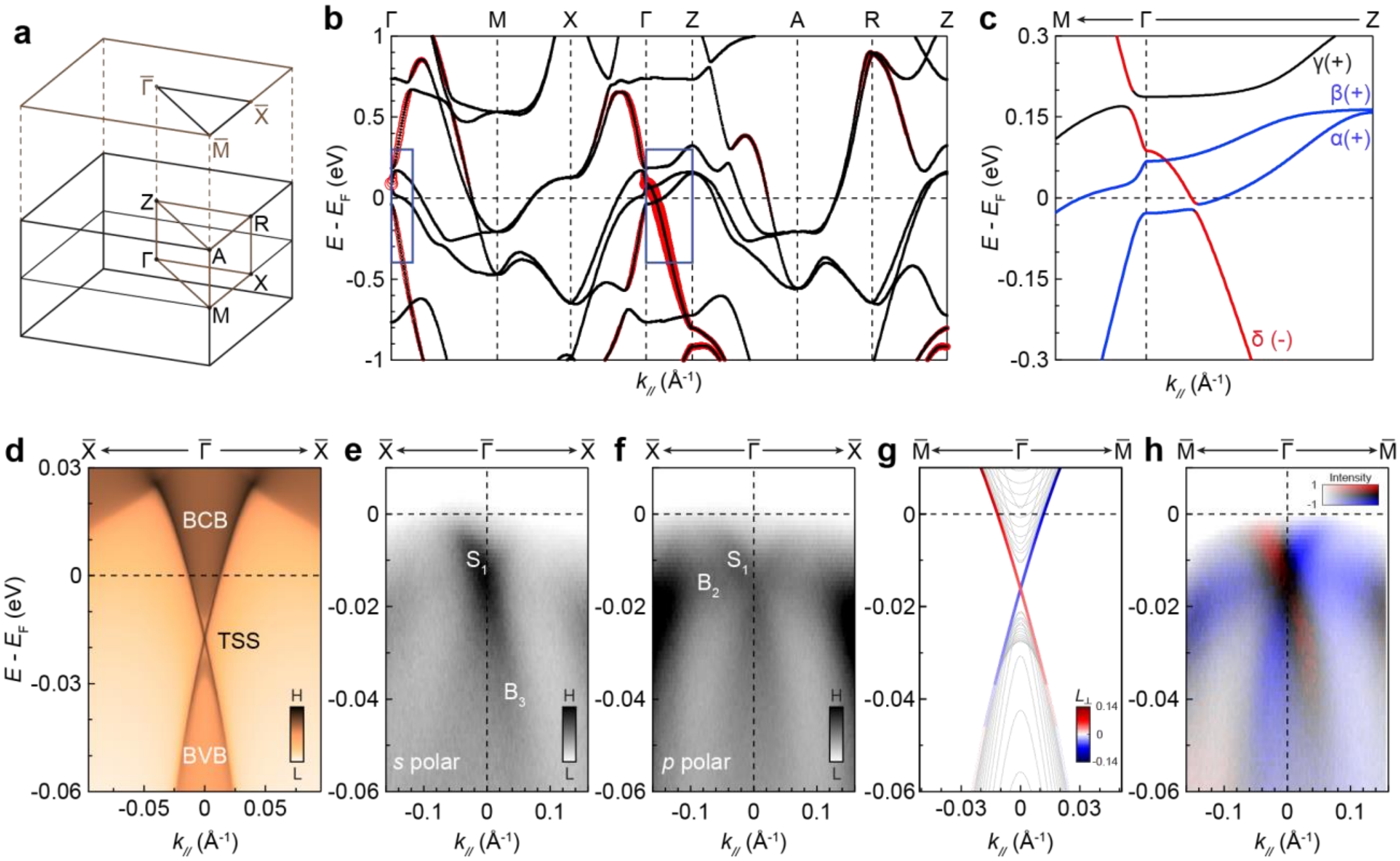

**Fig. 4| Topological superconductivity in the stoichiometric 20-UC FeTe film. a,** 3D BZ and projected surface BZ. **b,** Calculated band structure of stoichiometric FeTe with SOC along high-symmetry paths in the 3D BZ. The size of the red circles indicates the weight of the Te $p_z$ orbital. **c,** Enlarged view of the region outlined by the solid blue boxes in (**b**). The bands in (**c**) are labeled by their parity (±) at Γ, with red and blue indicating odd and even parity, respectively. The α and β bands are derived primarily from $d_{xz}/d_{yz}$ orbitals, whereas the δ band originates mainly from the $d_{xy}$ and Te $p_z$ orbitals. **d,** Band structure projected onto the (001) surface. **e,** Band dispersion measured at $T = 25$ K along the $\overline{\mathrm{X}}-\overline{\Gamma}-\overline{\mathrm{X}}$ direction using *s*-polarized photons with photon energy $hv = 6$ eV. **f,** Same as (**e**), but measured using *p*-polarized photons. **g,** Band dispersion from a slab calculation along the $\overline{\mathrm{M}}-\overline{\Gamma}-\overline{\mathrm{M}}$ direction, showing the atomic OAM texture of the TSS. **h,** Circular-dichroism ARPES band dispersion along the $\overline{\mathrm{M}}-\overline{\Gamma}-\overline{\mathrm{M}}$ direction. The circular-dichroism intensity is defined as the normalized difference $I_{\mathrm{CD}} = (I_{\mathrm{L}} - I_{\mathrm{R}})/(I_{\mathrm{L}} + I_{\mathrm{R}})$. The 2D color map encodes both the photoemission intensity and circular dichroism. Note that the DFT results use a different momentum scale to better match the ARPES spectra, with the calculated bands shifted by 74 meV towards higher binding energy to match the measured Dirac point position.

**References:**

1 Yan, Z. J., Wang, Z., Xia, B., Paolini, S., Chan, Y. T., Dihingia, N., Rong, H., Xiao, P., Halanayake, K. D., Song, J., Gowda, V., Hickey, D. R., Wu, W., Yu, J., Hirschfeld, P. J. & Chang, C. Z. Stoichiometric FeTe is a superconductor. *Nature* **652**, 342-348 (2026).

2 Fu, L. & Kane, C. L. Superconducting proximity effect and Majorana fermions at the surface of a topological insulator. *Phys. Rev. Lett.* **100**, 096407 (2008).

3 Zhang, P., Yaji, K., Hashimoto, T., Ota, Y., Kondo, T., Okazaki, K., Wang, Z., Wen, J., Gu, G. D., Ding, H. & Shin, S. Observation of topological superconductivity on the surface of an iron-based superconductor. *Science* **360**, 182-186 (2018).

4 Wang, D., Kong, L., Fan, P., Chen, H., Zhu, S., Liu, W., Cao, L., Sun, Y., Du, S., Schneeloch, J., Zhong, R., Gu, G., Fu, L., Ding, H. & Gao, H. J. Evidence for Majorana bound states in an iron-based superconductor. *Science* **362**, 333-335 (2018).

5 Wang, Z., Rodriguez, J. O., Jiao, L., Howard, S., Graham, M., Gu, G. D., Hughes, T. L., Morr, D. K. & Madhavan, V. Evidence for dispersing 1D Majorana channels in an iron-based superconductor. *Science* **367**, 104-108 (2020).

6 Zhu, S., Kong, L., Cao, L., Chen, H., Papaj, M., Du, S., Xing, Y., Liu, W., Wang, D., Shen, C., Yang, F., Schneeloch, J., Zhong, R., Gu, G., Fu, L., Zhang, Y. Y., Ding, H. & Gao, H. J. Nearly quantized conductance plateau of vortex zero mode in an iron-based superconductor. *Science* **367**, 189-192 (2020).

7 Stewart, G. R. Superconductivity in iron compounds. *Rev. Mod. Phys.* **83**, 1589-1652 (2011).

8 Dai, P. Antiferromagnetic order and spin dynamics in iron-based superconductors. *Rev. Mod. Phys.* **87**, 855-896 (2015).

9 Paglione, J. & Greene, R. L. High-temperature superconductivity in iron-based materials. *Nat. Phys.* **6**, 645-658 (2010).

10 Fernandes, R. M., Chubukov, A. V. & Schmalian, J. What drives nematic order in iron-based superconductors? *Nat. Phys.* **10**, 97-104 (2014).

11 Kamihara, Y., Watanabe, T., Hirano, M. & Hosono, H. Iron-Based Layered Superconductor

La[$O_{1-x}F_x$]FeAs ($x$ = 0.05-0.12) with $T_c$ = 26 K. *J. Am. Chem. Soc.* **130**, 3296-3297 (2008).

12 Chen, X. H., Wu, T., Wu, G., Liu, R. H., Chen, H. & Fang, D. F. Superconductivity at 43 K in $SmFeAsO_{1-x}F_x$. *Nature* **453**, 761-762 (2008).

13 Ren, Z.-A., Lu, W., Yang, J., Yi, W., Shen, X.-L., Cai, Z., Che, G.-C., Dong, X.-L., Sun, L.-L., Zhou, F. & Zhao, Z. Superconductivity at 55 K in Iron-Based F-Doped Layered Quaternary Compound Sm[$O_{1-x}F_x$]FeAs. *Chin. Phys. Lett.* **25**, 2215-2216 (2008).

14 Wang, Q.-Y., Li, Z., Zhang, W.-H., Zhang, Z.-C., Zhang, J.-S., Li, W., Ding, H., Ou, Y.-B., Deng, P., Chang, K., Wen, J., Song, C.-L., He, K., Jia, J.-F., Ji, S.-H., Wang, Y.-Y., Wang, L.-L., Chen, X., Ma, X.-C. & Xue, Q.-K. Interface-Induced High-Temperature Superconductivity in Single Unit-Cell FeSe Films on $SrTiO_3$. *Chin. Phys. Lett.* **29**, 037402 (2012).

15 Li, S., de la Cruz, C., Huang, Q., Chen, Y., Lynn, J. W., Hu, J., Huang, Y.-L., Hsu, F.-C., Yeh, K.-W., Wu, M.-K. & Dai, P. First-order magnetic and structural phase transitions in $Fe_{1+y}Se_xTe_{1-x}$. *Phys. Rev. B* **79**, 054503 (2009).

16 Ambolode, L. C. C., Okazaki, K., Horio, M., Suzuki, H., Liu, L., Ideta, S., Yoshida, T., Mikami, T., Kakeshita, T., Uchida, S., Ono, K., Kumigashira, H., Hashimoto, M., Lu, D. H., Shen, Z. X. & Fujimori, A. Dependence of electron correlation strength in $Fe_{1+y}Te_{1-x}Se_x$ on Se content. *Phys. Rev. B* **92**, 035104 (2015).

17 Bao, W., Qiu, Y., Huang, Q., Green, M. A., Zajdel, P., Fitzsimmons, M. R., Zhernenkov, M., Chang, S., Fang, M., Qian, B., Vehstedt, E. K., Yang, J., Pham, H. M., Spinu, L. & Mao, Z. Q. Tunable ($\delta\pi$, $\delta\pi$)-type antiferromagnetic order in α-Fe(Te,Se) superconductors. *Phys. Rev. Lett.* **102**, 247001 (2009).

18 Dong, C., Wang, H., Li, Z., Chen, J., Yuan, H. Q. & Fang, M. Revised phase diagram for the $FeTe_{1-x}Se_x$ system with fewer excess Fe atoms. *Phys. Rev. B* **84**, 224506 (2011).

19 Medvedev, S., McQueen, T. M., Troyan, I. A., Palasyuk, T., Eremets, M. I., Cava, R. J., Naghavi, S., Casper, F., Ksenofontov, V., Wortmann, G. & Felser, C. Electronic and magnetic phase diagram of $\beta$-$Fe_{1.01}Se$ with superconductivity at 36.7 K under pressure. *Nat. Mater.* **8**,

630-633 (2009).

20 Wang, Z., Zhang, P., Xu, G., Zeng, L. K., Miao, H., Xu, X., Qian, T., Weng, H., Richard, P., Fedorov, A. V., Ding, H., Dai, X. & Fang, Z. Topological nature of the $FeSe_{0.5}Te_{0.5}$ superconductor. *Phys. Rev. B* **92**, 115119 (2015).

21 Li, Y. F., Chen, S. D., García-Díez, M., Iraola, M. I., Pfau, H., Zhu, Y. L., Mao, Z. Q., Chen, T., Yi, M., Dai, P. C., Sobota, J. A., Hashimoto, M., Vergniory, M. G., Lu, D. H. & Shen, Z. X. Orbital Ingredients and Persistent Dirac Surface State for the Topological Band Structure in $FeTe_{0.55}Se_{0.45}$. *Phys. Rev. X* **14**, 021043 (2024).

22 Lin, H., Jacobs, C. L., Yan, C., Nolan, G. M., Berruto, G., Singleton, P., Nguyen, K. D., Bai, Y., Gao, Q., Wu, X., Liu, C. X., Yan, G., Choi, S., Liu, C., Guisinger, N. P., Huang, P. Y., Mandal, S. & Yang, S. A topological superconductor tuned by electronic correlations. *Nat. Commun.* **17**, 1188 (2025).

23 Kim, Y., Yoo, J., Kim, S., Hahn, S., Tanaka, K., Yu, L., Kim, M. & Kim, C. Fragility of Topology under Electronic Correlations in Iron Chalcogenides. *Phys. Rev. Lett.* **136**, 196502 (2026).

24 Ma, F., Ji, W., Hu, J., Lu, Z. Y. & Xiang, T. First-principles calculations of the electronic structure of tetragonal α-FeTe and α-FeSe crystals: evidence for a bicollinear antiferromagnetic order. *Phys. Rev. Lett.* **102**, 177003 (2009).

25 Rodriguez, E. E., Stock, C., Zajdel, P., Krycka, K. L., Majkrzak, C. F., Zavalij, P. & Green, M. A. Magnetic-crystallographic phase diagram of the superconducting parent compound $Fe_{1+x}Te$. *Phys. Rev. B* **84**, 064403 (2011).

26 Ducatman, S., Fernandes, R. M. & Perkins, N. B. Theory of the evolution of magnetic order in $Fe_{1+y}Te$ compounds with increasing interstitial iron. *Phys. Rev. B* **90**, 165123 (2014).

27 de la Cruz, C., Huang, Q., Lynn, J. W., Li, J., Ratcliff II, W., Zarestky, J. L., Mook, H. A., Chen, G. F., Luo, J. L., Wang, N. L. & Dai, P. Magnetic order close to superconductivity in the iron-based layered $LaO_{1-x}F_xFeAs$ systems. *Nature* **453**, 899-902 (2008).

28 Huang, Q., Qiu, Y., Bao, W., Green, M. A., Lynn, J. W., Gasparovic, Y. C., Wu, T., Wu, G. &

Chen, X. H. Neutron-diffraction measurements of magnetic order and a structural transition in the parent $BaFe_2As_2$ compound of FeAs-based high-temperature superconductors. *Phys. Rev. Lett.* **101**, 257003 (2008).

29 Xia, Y., Qian, D., Wray, L., Hsieh, D., Chen, G. F., Luo, J. L., Wang, N. L. & Hasan, M. Z. Fermi surface topology and low-lying quasiparticle dynamics of parent $Fe_{1+x}Te/Se$ superconductor. *Phys. Rev. Lett.* **103**, 037002 (2009).

30 Zhang, Y., Chen, F., He, C., Yang, L. X., Xie, B. P., Xie, Y. L., Chen, X. H., Fang, M., Arita, M., Shimada, K., Namatame, H., Taniguchi, M., Hu, J. P. & Feng, D. L. Strong correlations and spin-density-wave phase induced by a massive spectral weight redistribution in α-$Fe_{1.06}Te$. *Phys. Rev. B* **82**, 165113 (2010).

31 Liu, Z. K., He, R. H., Lu, D. H., Yi, M., Chen, Y. L., Hashimoto, M., Moore, R. G., Mo, S. K., Nowadnick, E. A., Hu, J., Liu, T. J., Mao, Z. Q., Devereaux, T. P., Hussain, Z. & Shen, Z. X. Measurement of coherent polarons in the strongly coupled antiferromagnetically ordered iron-chalcogenide $Fe_{1.02}Te$ using angle-resolved photoemission spectroscopy. *Phys. Rev. Lett.* **110**, 037003 (2013).

32 Kim, Y., Kim, M. S., Kim, D., Kim, M., Kim, M., Cheng, C. M., Choi, J., Jung, S., Lu, D., Kim, J. H., Cho, S., Song, D., Oh, D., Yu, L., Choi, Y. J., Kim, H. D., Han, J. H., Jo, Y., Shim, J. H., Seo, J., Huh, S. & Kim, C. Kondo interaction in FeTe and its potential role in the magnetic order. *Nat. Commun.* **14**, 4145 (2023).

33 Lin, P. H., Texier, Y., Taleb-Ibrahimi, A., Le Fèvre, P., Bertran, F., Giannini, E., Grioni, M. & Brouet, V. Nature of the bad metallic behavior of $Fe_{1.06}Te$ inferred from its evolution in the magnetic state. *Phys. Rev. Lett.* **111**, 217002 (2013).

34 Xu, H., Jiang, J., Gai, X., Cao, R. Q., Chen, K., Man, X. X., Lin, H., Deng, P., He, K., Liu, K., Zhao, D., Lu, Z. Y., Chang, K. & Liu, C. Reversible Tuning of Magnetic Order and Intrinsic Superconductivity in Strained FeTe Films via Stoichiometry Control. *ACS Nano* **20**, 16426-16434 (2026).

35 Lifshitz, I. M. Anomalies of electron characteristics of a metal in the high pressure region.

*Sov. Phys. JETP* **11**, 1130-1135 (1960).

36 Kuroki, K., Higashida, T. & Arita, R. High-$T_c$ superconductivity due to coexisting wide and narrow bands: A fluctuation exchange study of the Hubbard ladder as a test case. *Phys. Rev. B* **72**, 212509 (2005).

37 Chen, X., Maiti, S., Linscheid, A. & Hirschfeld, P. J. Electron pairing in the presence of incipient bands in iron-based superconductors. *Phys. Rev. B* **92**, 224514 (2015).

38 Linscheid, A., Maiti, S., Wang, Y., Johnston, S. & Hirschfeld, P. J. High $T_c$ via Spin Fluctuations from Incipient Bands: Application to Monolayers and Intercalates of FeSe. *Phys. Rev. Lett.* **117**, 077003 (2016).

39 Huang, J., Yu, R., Xu, Z., Zhu, J.-X., Oh, J. S., Jiang, Q., Wang, M., Wu, H., Chen, T., Denlinger, J. D., Mo, S.-K., Hashimoto, M., Michiardi, M., Pedersen, T. M., Gorovikov, S., Zhdanovich, S., Damascelli, A., Gu, G., Dai, P., Chu, J.-H., Lu, D., Si, Q., Birgeneau, R. J. & Yi, M. Correlation-driven electronic reconstruction in $FeTe_{1-x}Se_x$. *Commun. Phys.* **5**, 29 (2022).

40 Armitage, N. P., Ronning, F., Lu, D. H., Kim, C., Damascelli, A., Shen, K. M., Feng, D. L., Eisaki, H., Shen, Z. X., Mang, P. K., Kaneko, N., Greven, M., Onose, Y., Taguchi, Y. & Tokura, Y. Doping dependence of an n-type cuprate superconductor investigated by angle-resolved photoemission spectroscopy. *Phys. Rev. Lett.* **88**, 257001 (2002).

41 Song, D., Han, G., Kyung, W., Seo, J., Cho, S., Kim, B. S., Arita, M., Shimada, K., Namatame, H., Taniguchi, M., Yoshida, Y., Eisaki, H., Park, S. R. & Kim, C. Electron Number-Based Phase Diagram of $Pr_{1-x}LaCe_xCuO_{4-\delta}$ and Possible Absence of Disparity between Electron- and Hole-Doped Cuprate Phase Diagrams. *Phys. Rev. Lett.* **118**, 137001 (2017).

42 Chen, F., Zhou, B., Zhang, Y., Wei, J., Ou, H.-W., Zhao, J.-F., He, C., Ge, Q.-Q., Arita, M., Shimada, K., Namatame, H., Taniguchi, M., Lu, Z.-Y., Hu, J., Cui, X.-Y. & Feng, D. L. Electronic structure of $Fe_{1.04}Te_{0.66}Se_{0.34}$. *Phys. Rev. B* **81**, 014526 (2010).

43 Lubashevsky, Y., Lahoud, E., Chashka, K., Podolsky, D. & Kanigel, A. Shallow pockets and very strong coupling superconductivity in $FeSe_xTe_{1-x}$. *Nat. Phys.* **8**, 309-312 (2012).

44 Watson, M. D., Yamashita, T., Kasahara, S., Knafo, W., Nardone, M., Beard, J., Hardy, F., McCollam, A., Narayanan, A., Blake, S. F., Wolf, T., Haghighirad, A. A., Meingast, C., Schofield, A. J., Lohneysen, H., Matsuda, Y., Coldea, A. I. & Shibauchi, T. Dichotomy between the Hole and Electron Behavior in Multiband Superconductor FeSe Probed by Ultrahigh Magnetic Fields. *Phys. Rev. Lett.* **115**, 027006 (2015).

45 Yoshikawa, N., Takayama, M., Shikama, N., Ishikawa, T., Nabeshima, F., Maeda, A. & Shimano, R. Charge carrier dynamics of FeSe thin film investigated by terahertz magneto-optical spectroscopy. *Phys. Rev. B* **100**, 035110 (2019).

46 Zhang, P., Wang, Z., Wu, X., Yaji, K., Ishida, Y., Kohama, Y., Dai, G., Sun, Y., Bareille, C., Kuroda, K., Kondo, T., Okazaki, K., Kindo, K., Wang, X., Jin, C., Hu, J., Thomale, R., Sumida, K., Wu, S., Miyamoto, K., Okuda, T., Ding, H., Gu, G. D., Tamegai, T., Kawakami, T., Sato, M. & Shin, S. Multiple topological states in iron-based superconductors. *Nat. Phys.* **15**, 41-47 (2019).

47 Kim, M., Choi, S., Brito, W. H. & Kotliar, G. Orbital-Selective Mott Transition Effects and Nontrivial Topology of Iron Chalcogenide. *Phys. Rev. Lett.* **132**, 136504 (2024).

48 Hsieh, D., Xia, Y., Qian, D., Wray, L., Dil, J. H., Meier, F., Osterwalder, J., Patthey, L., Checkelsky, J. G., Ong, N. P., Fedorov, A. V., Lin, H., Bansil, A., Grauer, D., Hor, Y. S., Cava, R. J. & Hasan, M. Z. A tunable topological insulator in the spin helical Dirac transport regime. *Nature* **460**, 1101-1105 (2009).

49 Hsieh, D., Xia, Y., Wray, L., Qian, D., Pal, A., Dil, J. H., Osterwalder, J., Meier, F., Bihlmayer, G., Kane, C. L., Hor, Y. S., Cava, R. J. & Hasan, M. Z. Observation of unconventional quantum spin textures in topological insulators. *Science* **323**, 919-922 (2009).

50 Pan, Z. H., Vescovo, E., Fedorov, A. V., Gardner, D., Lee, Y. S., Chu, S., Gu, G. D. & Valla, T. Electronic structure of the topological insulator $Bi_2Se_3$ using angle-resolved photoemission spectroscopy: evidence for a nearly full surface spin polarization. *Phys. Rev. Lett.* **106**, 257004 (2011).

51 Souma, S., Kosaka, K., Sato, T., Komatsu, M., Takayama, A., Takahashi, T., Kriener, M., Segawa, K. & Ando, Y. Direct measurement of the out-of-plane spin texture in the Dirac-cone surface state of a topological insulator. *Phys. Rev. Lett.* **106**, 216803 (2011).

52 Sidilkover, I., Yen, Y., D'Souza, S. W., Schusser, J., Pulkkinen, A., Rotundu, C. R., Hashimoto, M., Liu, D., Shen, Z.-X., Minár, J., Schüler, M., Soifer, H. & Sobota, J. A. Reexamining circular dichroism in photoemission from a topological insulator. *Phys. Rev. Res.* **7**, 033027 (2025).

53 Kuroki, K., Onari, S., Arita, R., Usui, H., Tanaka, Y., Kontani, H. & Aoki, H. Unconventional pairing originating from the disconnected Fermi surfaces of superconducting $LaFeAsO_{1-x}F_x$. *Phys. Rev. Lett.* **101**, 087004 (2008).

54 Mazin, I. I., Singh, D. J., Johannes, M. D. & Du, M. H. Unconventional superconductivity with a sign reversal in the order parameter of $LaFeAsO_{1-x}F_x$. *Phys. Rev. Lett.* **101**, 057003 (2008).

55 Terashima, K., Sekiba, Y., Bowen, J. H., Nakayama, K., Kawahara, T., Sato, T., Richard, P., Xu, Y. M., Li, L. J., Cao, G. H., Xu, Z. A., Ding, H. & Takahashi, T. Fermi surface nesting induced strong pairing in iron-based superconductors. *Proc. Natl. Acad. Sci. U. S. A.* **106**, 7330-7333 (2009).

56 Hirschfeld, P. J., Korshunov, M. M. & Mazin, I. I. Gap symmetry and structure of Fe-based superconductors. *Rep. Prog. Phys.* **74**, 124508 (2011).

57 Chubukov, A. Pairing Mechanism in Fe-Based Superconductors. *Annu. Rev. Condens. Matter Phys.* **3**, 57-92 (2012).

58 Li, C., Yan, Z.-J. Y., Ge, Y., Wang, Z., Xia, B., Paolini, S., Xiao, P., Lai, L.-K., Song, J., Kaczmarek, A. R., Min, L., Yasuda, K., Hirschfeld, P. J., Yu, J., Chang, C.-Z. & Nowack, K. C. Signatures of nodal superconductivity in stoichiometric FeTe. *arXiv:2609.08116* (2026).

59 Hirschfeld, P. J. Using gap symmetry and structure to reveal the pairing mechanism in Fe-based superconductors. *C. R. Physique* **17**, 197-231 (2015).

60 Hua, Y., Yang, W.-l., Miao, J.-J., Xu, H. & Yue, C. Competing Extended-s- and d-Wave

Pairing from Distinct Spin-Fluctuation Channels in Stoichiometric FeTe. *arXiv:2608.23467* (2026).

61 Yan, C., Green, E., Fukumori, R., Protic, N., Lee, S. H., Fernandez-Mulligan, S., Raja, R., Erdakos, R., Mao, Z. & Yang, S. An integrated quantum material testbed with multi-resolution photoemission spectroscopy. *Rev. Sci. Instrum.* **92**, 113907 (2021).

62 Horcas, I., Fernandez, R., Gomez-Rodriguez, J. M., Colchero, J., Gomez-Herrero, J. & Baro, A. M. WSXM: a software for scanning probe microscopy and a tool for nanotechnology. *Rev. Sci. Instrum.* **78**, 013705 (2007).

63 Kresse, G. & Hafner, J. Ab initio molecular dynamics for liquid metals. *Phys. Rev. B* **47**, 558-561 (1993).

64 Kresse, G. & Furthmüller, J. Efficiency of ab-initio total energy calculations for metals and semiconductors using a plane-wave basis set. *Comput. Mater. Sci.* **6**, 15-50 (1996).

65 Kresse, G. & Hafner, J. Ab initio molecular-dynamics simulation of the liquid-metal-amorphous-semiconductor transition in germanium. *Phys. Rev. B* **49**, 14251-14269 (1994).

66 Kresse, G. & Furthmuller, J. Efficient iterative schemes for ab initio total-energy calculations using a plane-wave basis set. *Phys. Rev. B* **54**, 11169-11186 (1996).

67 Kresse, G. & Joubert, D. From ultrasoft pseudopotentials to the projector augmented-wave method. *Phys. Rev. B* **59**, 1758-1775 (1999).

68 Perdew, J. P., Burke, K. & Ernzerhof, M. Generalized Gradient Approximation Made Simple. *Phys. Rev. Lett.* **77**, 3865-3868 (1996).

69 Mostofi, A. A., Yates, J. R., Pizzi, G., Lee, Y.-S., Souza, I., Vanderbilt, D. & Marzari, N. An updated version of wannier90: A tool for obtaining maximally-localised Wannier functions. *Comput. Phys. Commun.* **185**, 2309-2310 (2014).

70 Sancho, M. P. L., Sancho, J. M. L., Sancho, J. M. L. & Rubio, J. Highly convergent schemes for the calculation of bulk and surface Green functions. *J. Phys. F: Met. Phys.* **15**, 851-858 (1985).

71 Yu, R., Qi, X. L., Bernevig, A., Fang, Z. & Dai, X. Equivalent expression of $Z_2$ topological

invariant for band insulators using the non-Abelian Berry connection. *Phys. Rev. B* **84** (2011).

72 Kane, C. L. & Mele, E. J. $Z_2$ Topological Order and the Quantum Spin Hall Effect. *Phys. Rev. Lett.* **95**, 146802 (2005).

73 Haule, K., Yee, C.-H. & Kim, K. Dynamical mean-field theory within the full-potential methods: Electronic structure of $CeIrIn_5$, $CeCoIn_5$, and$CeRhIn_5$. *Phys. Rev. B* **81**, 195107 (2010).

74 Kotliar, G., Savrasov, S. Y., Haule, K., Oudovenko, V. S., Parcollet, O. & Marianetti, C. A. Electronic structure calculations with dynamical mean-field theory. *Rev. Mod. Phys.* **78**, 865-951 (2006).

75 Haule, K. Exact Double Counting in Combining the Dynamical Mean Field Theory and the Density Functional Theory. *Phys. Rev. Lett.* **115**, 196403 (2015).

76 Haule, K. & Birol, T. Free Energy from Stationary Implementation of the DFT+DMFT Functional. *Phys. Rev. Lett.* **115**, 256402 (2015).

77 Haule, K. eDMFT: Density Functional Theory + Embedded Dynamical Mean Field Theory.

78 Blaha, P., Schwarz, K., Tran, F., Laskowski, R., Madsen, G. K. H. & Marks, L. D. WIEN2k: An APW+lo program for calculating the properties of solids. *J. Chem. Phys.* **152**, 074101 (2020).

79 Haule, K. Quantum Monte Carlo impurity solver for cluster dynamical mean-field theory and electronic structure calculations with adjustable cluster base. *Phys. Rev. B* **75**, 155113 (2007).

80 Han, M. J. & Savrasov, S. Y. Doping Driven (π, 0) Nesting and Magnetic Properties of $Fe_{1+x}Te$ Superconductors. *Phys. Rev. Lett.* **103**, 067001 (2009).

81 Zhang, L., Singh, D. J. & Du, M. H. Density functional study of excess Fe in $Fe_{1+x}Te$: Magnetism and doping. *Phys. Rev. B* **79**, 012506 (2009).

82 Han, M. J. & Savrasov, S. Y. Han and Savrasov Reply. *Phys. Rev. Lett.* **104**, 099702 (2010).

83 Singh, P. P. Comment on “Doping Driven (π, 0) Nesting and Magnetic Properties of $Fe_{1+x}Te$ Superconductors”. *Phys. Rev. Lett.* **104**, 099701 (2010).

84 Mandal, S., Zhang, P., Ismail-Beigi, S. & Haule, K. How Correlated is the $FeSe/SrTiO_3$ System? *Phys. Rev. Lett.* **119**, 067004 (2017).

85 Yin, Z. P., Haule, K. & Kotliar, G. Kinetic frustration and the nature of the magnetic and paramagnetic states in iron pnictides and iron chalcogenides. *Nat. Mater.* **10**, 932-935 (2011).